\documentclass[twocolumn,showpacs,floats,floatfix,superscriptaddress,aps,pra]{revtex4}

\usepackage{amsfonts,amssymb,amsmath}
\usepackage{color,calc}
\usepackage[dvips]{graphicx}
\usepackage{bm}

\def\be{\begin{equation}}
\def\ee{\end{equation}}
\def\bea{\begin{eqnarray}}
\def\eea{\end{eqnarray}}
\def\bse{\begin{subequations}}
\def\ese{\end{subequations}}
\def\bt{\begin{tabular}}
\def\et{\end{tabular}}
\def\bc{\begin{center}}
\def\ec{\end{center}}

\def\ups{\vert\!\!\uparrow\rangle}
\def\downs{\vert\!\!\downarrow\rangle}

\def\ket#1{\vert #1 \rangle}
\def\bra#1{\langle #1 \vert}

\def\etal{\textit{et al.}}

\def\Bosc{B^\prime}

\def\hes{A}

\def\id{\mathbf{1}}

\def\Nd{N}
\def\id{\mathbf{1}}
\def\CP{C}
\def\ph{T}
\def\cph{C}
\def\H{H}
\def\HI{\H_\text{I}}
\def\HS{\H_\text{S}}
\def\U{U}

\def\r{\rho}
\def\cm{\text{c.m.}}
\def\tsum{\displaystyle{\sum}}

\begin{document}

\title{Simplified implementation of the quantum Fourier transform with Ising-type Hamiltonians: Example with ion traps}
\author{Svetoslav S. Ivanov}
\affiliation{Department of Physics, Sofia University, 5 James Bourchier Blvd, 1164 Sofia, Bulgaria}
\author{Michael Johanning}
\affiliation{Department of Physics, University of Siegen, Walter-Flex-Strasse 3, Siegen NRW 57072, Germany}
\author{Christof Wunderlich}
\affiliation{Department of Physics, University of Siegen, Walter-Flex-Strasse 3, Siegen NRW 57072, Germany}

\date{\today}

\begin{abstract}
We propose a simplified mathematical construction of the quantum Fourier transform which is suited for systems
described by Ising-type Hamiltonians.
By contrast to the standard Cooley-Tuckey scheme, which prescribes sequences of CPHASE gates,
our implementation is based on one-qubit gates and a free evolution process.
We also show how to obtain a quadratic speed-up by applying the conditional interactions simultaneously.
Thus rather than O($N^2$) our implementation time scales as O($N$).
Finally, we show a realization of our method with homogeneous microwave driven ion traps in a magnetic field with gradient.
\end{abstract}


\maketitle

\section{Introduction}\label{Sec-introduction}

The quantum discrete Fourier transform (QFT) is the fundamental ingredient of many principal quantum algorithms,
known to provide substantial speed-up over their classical counterparts \cite{Chuang,Jozsa,Bowden},
including Shor's order finding, factorization and the discrete logarithm \cite{Shor}, Deutsch's algorithm \cite{Deutsch} and
phase estimation \cite{Kitaev}. Being at the heart of quantum computation QFT
has become a textbook example of the potential superiority of quantum over classical computers
and thus it has inspired extensive research aiming at the implementation
of this transformation in different physical systems.

Experimental demonstrations include the composition of two- and three-qubit QFT in nuclear magnetic resonance (NMR) \cite{Cory2001,ChuangQFT2001,ChuangPRL2000,PhaseEstimation2002},
neutral molecules \cite{neutralQFT2010}, superconducting qubits \cite{SC-QFT2011},
and semiclassical realizations with trapped ions \cite{Wineland2005} and photonic qubits \cite{photonicQFT}.
These realizations use predominantly the Cooley-Tuckey algorithm \cite{CT1965}, which gives an efficient implementation of QFT
with a polynomial number O($N^2$) of conditional gates, with $N$ being the number of qubits.
Despite the mathematical efficiency of this algorithm, however, the attempts to scale the number of qubits have quickly come across a serious obstacle:
the number of gates and the ensuing physical operations with limited precision still grow rapidly with the size of the problem.
They reach prohibitively large values before approaching even moderate computational scales.
Therefore one must seek for more realistic implementation models which require less resources.
One way to go is by decomposing QFT directly into more favourable gates, which can be realized naturally in the particular physical system.

In this work we present a mathematical construction of QFT, which is a modification of the Cooley-Tuckey scheme.
Our circuit is suited for a broad range of systems described by Ising-type Hamiltonians of the form 
\be
\label{Hamiltonian}
\HI = -\frac{\hbar}{2}\sum_{k<l=1}^{\Nd}J_{kl}\sigma^{(k)}_{z}\sigma^{(l)}_{z},
\ee
where $\sigma^{(k)}_{z}$ represent the spins and $J_{kl}$ represent coupling strengths.
Examples for such systems are nuclear magnetic resonance \cite{ChuangQFT2001}, ion traps \cite{Wunderlich}, spin chains.
Our proposal has two principal advantages.
First, the sequences of conditional gates, which are the major stumbling block to QFT,
are now carried out by simple time evolutions driven by the Hamiltonian \eqref{Hamiltonian} over certain time intervals.
Thus in our implementation QFT is realized with a series of one-qubit gates applied at particular times.
This implies that, second, the number of physical interactions with the system is reduced.
We also show how to obtain quadratically faster QFT by applying the conditional interactions simultaneously.

We include an explicit  discussion of  the physical realization of the proposed scheme with trapped ions.

\section{Quantum Fourier Transform: Mathematical construction}\label{Sec-theory}

The $\Nd$-dimensional QFT is defined through its action on the computational basis states $\ket{0}$, $\ket{1}$, $\ket{2}$, $\ldots$, $\ket{\Nd-1}$ by
\be
F_{\Nd}\ket{n} = \frac{1}{\sqrt{\Nd}} \sum_{k=0}^{\Nd-1} e^{2\pi i n k/\Nd}\ket{k}.
\ee
It is a unitary transformation, which maps each state $\ket{n}$ into an equal superposition of all states each imprinted with a particular phase.

\begin{figure}[tbf]
\begin{center}
\includegraphics[angle=0,width=1\columnwidth]{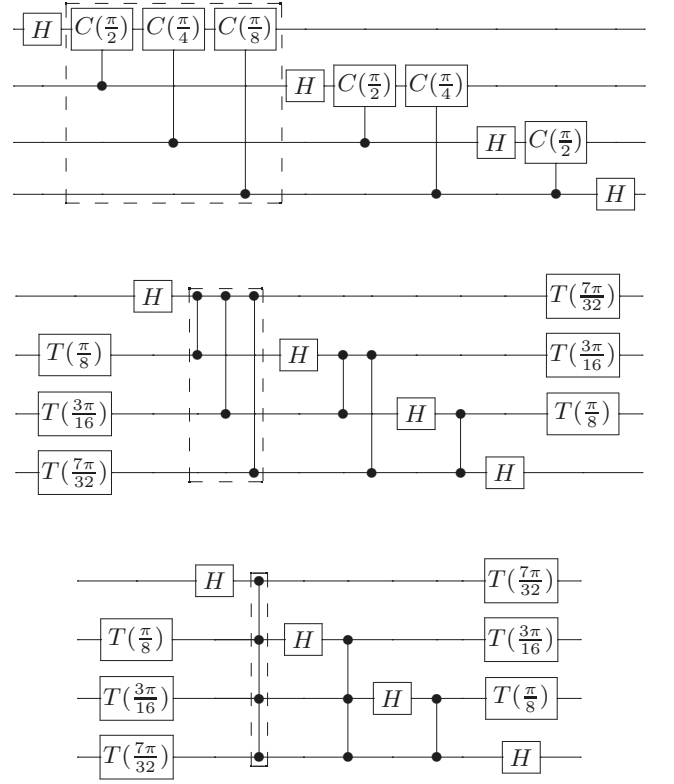}
\end{center}
\caption{
(top) Decomposition of the quantum Fourier transform using standard gates from the circuit model \cite{Shor,Chuang}. The example shown is for $\Nd=4$ qubits. Here $H$ is the Hadamard gate and $\cph(\phi_k)$ are control phase gates (see text for details).
(middle) Decomposition of the same transform based on the gate $U(\phi)$, designated with two joined circles. Note that $U(\phi)$ can be achieved with simple free evolution, driven by the Hamiltonian \eqref{Hamiltonian}. $T(\phi)$ is a one-qubit phase gate. See Eqs. \eqref{gates}.
In both cases not shown are swap gates at the end of the circuit, which exchange qubits $1 \leftrightarrow 4$ and $2 \leftrightarrow 3$.
(bottom) Simultaneous conditional evolution is employed.
}
\label{fig1}
\end{figure}

A circuit that implements QFT is shown in Fig. \ref{fig1} (top). It is based on the Cooley-Tuckey algorithm and can be found in Refs.~\cite{Chuang} and \cite{CoppersmithDeutsch94}.
Two gates from the elementary gate set of the standard computational model are used --
the single-qubit Hadamard gate
\be
H_k=\tfrac{1}{\sqrt{2}} \left[\begin{array}{cc} 1 & 1 \\ 1 & -1 \end{array}\right],
\ee
acting on qubit $k$ and the two-qubit conditional phase gate (CPHASE gate)
\be
\cph_{kl}(\phi)=\left[
    \begin{array}{cccc}
      1 & 0 & 0 & 0 \\
      0 & 1 & 0 & 0 \\
      0 & 0 & 1 & 0 \\
      0 & 0 & 0 & e^{i\phi} \\
    \end{array}
  \right],
\ee
acting on qubits $k$ and $l$, where
\be
\label{phases}
\phi=\pi/2^{l-k}.
\ee
This circuit offers an exponentially faster implementation (with $O(\Nd^2)$ gates) compared to the best known classical algorithm for fast Fourier transform (with $O(\Nd 2^\Nd)$ gates).
Despite this formal superiority, however, this implementation has proved difficult to scale due to the sequences of the CPHASE gates.
Hence other mathematical constructions of QFT are desirable, which use more favourable gates for the particular physical system rather than the CPHASE gates.

Based on the Cooley-Tuckey scheme, in the following we give a mathematical construction of QFT, which is adapted for a class of systems described by the Hamiltonian \eqref{Hamiltonian}.
This Hamiltonian realizes long-range spin-spin coupling and, since it is time-independent, the evolution matrix is $U=\text{exp}\left(-i \HI t/\hbar\right)$, or
\be
\label{propagator}
U=\text{exp}\left(i\sum_{k<l=1}^N\phi_{kl}\sigma_z^{(k)}\sigma_z^{(l)}\right).
\ee
Below we will also use the notation $U(t)=\text{exp}\left[i t\sum_{k<l=1}^N J_{kl}\sigma_z^{(k)}\sigma_z^{(l)}\right]$ when we want to highlight the dependence of $U$ on time.
Note that $U$ yields control-phase shifts between all pairs of spins,
$U = \prod_{k<l=1}^{N}U_{kl}(\phi_{kl})$,
where $U_{kl}(\phi_{kl}) = \text{exp}\left(i\phi_{kl}\sigma_z^{(k)}\sigma_z^{(l)}\right)$. The corresponding phase is $\phi_{kl} = J_{kl}t/2$. Thus $U_{kl}(\phi_{kl})$, which will substitute the CPHASE gate, is merely delivered by free evolution after time $t=2\phi_{kl}/J_{kl}$. Thus, the required free evolution time might be different for each $\phi_{kl}$; however, if the coupling $J_{kl}$ can be tailored such, that all $t$ are identical, all phases $\phi_{kl}$ can be acquired in a single step. In NMR, there exist proposals for the implementation of the two-qubit CNOT gate and multi-qubit gates, based on always-on interactions (by refocussing other interactions, and thus ``wasting them'') \cite{Schulte2005}.

Our gate set comprises a rotation $R_k(\phi)$, a phase gate $\ph_k(\phi)$, and the entangling gate $U_{kl}(\phi)$. The indices indicate qubit numbers.
We will also use implementations, where the phase gate is incorporated in the rotation to form $R(\theta,\phi)$ according to the rule $R(\theta,\phi) = \ph(\phi/2)R(\theta)\ph(-\phi/2)$.
Such phased rotation can be obtained with a simple phase shift $\phi$ in the driving field implementing $R(\theta)$.
We have
\bse
\label{gates}
\begin{align}
  R_k(\theta,\phi) &= e^{-i \frac{\theta}{2} (\sigma^{(k)}_{x} \cos\phi + \sigma^{(k)}_{y} \sin\phi)}, \\
\ph_k(\phi) &= e^{-i \phi \sigma^{(k)}_{z}}, \\
U_{kl}(\phi)     &= e^{i \phi \, \sigma^{(k)}_{z} \sigma^{(l)}_{z}}, \label{U}
\end{align}
\ese
where 
$R_k(\theta) = R_k(\theta,0)$.
Sometimes we will drop the phase dependence in the gates, e.g. $U_{kl}(\phi_{kl})$ will be written more concisely as $U_{kl}$.

We derive our implementation from the circuit shown in Fig. \ref{fig1} (top), where we replace the CPHASE gates with free evolution $U(\phi)$ and phase gates $\ph(\phi)$.
To this end, we note that a CPHASE gate $\cph_{kl}(\phi)$ acting on qubits $k$ and $l$ can be written as
$\cph_{kl}(\phi)=e^{i\phi\ket{11}_{kl}\bra{11}_{kl}}\equiv e^{i\phi(\ket{1}\bra{1})_{k}\otimes(\ket{1}\bra{1})_{l}}$ and,
using the identity $\ket{1}\bra{1}=(\id-\sigma_z)/2$, we get $\cph_{kl}(\phi)=e^{i\frac{\phi}{4}(\id-\sigma^{(k)}_{z})(\id-\sigma^{(l)}_{z})}$.
After expanding the product we arrive at
\be
\cph_{kl}(\phi)=e^{i\frac{\phi}{4}}\ph_k(\tfrac{\phi}{4})\ph_l(\tfrac{\phi}{4})U_{kl}(\tfrac{\phi}{4}),
\ee
where all gates commute.
Hence, up to an unimportant global phase, which will be omitted in what follows, a CPHASE gate can be achieved with two phase gates and the free evolution gate.

Next we proceed with reconstructing the gate sequence from Fig. \ref{fig1} (top), where our aim is to obtain a circuit involving only free evolution gates and a minimum number of single-qubit gates. We note that the gate series, encircled with a dash line, can be expressed as
\begin{align}
\label{series}
&\CP_{12}(\phi_2)\CP_{13}(\phi_3)\CP_{14}(\phi_4)= \notag\\
&\exp\left[\tfrac{i}{4}\left(\phi_{2} \sigma^{(1)}_{z}\sigma^{(2)}_{z}+\phi_{3} \sigma^{(1)}_{z}\sigma^{(3)}_{z}+\phi_{4} \sigma^{(1)}_{z}\sigma^{(4)}_{z} \right)\right] \\
&\ph_1\left(\tfrac{\phi_1}{4}\right)\ph_2\left(\tfrac{\phi_2}{4}\right)\ph_3\left(\tfrac{\phi_3}{4}\right)\ph_4\left(\tfrac{\phi_4}{4}\right), \notag
\end{align}
where $\phi_1 = \phi_2 + \phi_3 + \phi_4$ or $\phi_1 = \sum_{l=2}^\Nd \phi_l$ for any $\Nd$. If we substitute the phases from Eq. \eqref{phases}, we get
\bse
\begin{align}
&\phi_1 = \tfrac{\pi}{2}(1-2^{-\Nd+1}),\\
&\phi_l = \pi/2^{l-1} \text{ for } l>1.
\end{align}
\ese
We concatenate such sequences of CPHASE gates with Hadamard gates, according to Fig. \ref{fig1} (top), and then we rearrange the so-obtained circuit such that
all phase gates to the left (right) of the Hadamard gates are combined in a single phase gate placed at the beginning (end) of the circuit.
Thus we reconstruct the entire circuit in the form shown in Fig. \ref{fig1} (middle).
Therefore, the Fourier transform over $\Nd=4$ qubits can be achieved with the following gate sequence (in operator form, to be read from right to left)
\begin{align}
\label{QFT4series}
T_\text{F} ~H_4 ~e^{i \tfrac{\pi}{4}\left(\tfrac{1}{2}\sigma^{(3)}_{z}\sigma^{(4)}_{z}\right)} H_3 ~e^{i \tfrac{\pi}{4}\left(\tfrac{1}{2}\sigma^{(2)}_{z}\sigma^{(3)}_{z}+\tfrac{1}{4}\sigma^{(2)}_{z}\sigma^{(4)}_{z}\right)} \times \notag\\
\times  H_2 ~e^{i \tfrac{\pi}{4}\left(\tfrac{1}{2}\sigma^{(1)}_{z}\sigma^{(2)}_{z}+\tfrac{1}{4}\sigma^{(1)}_{z}\sigma^{(3)}_{z}+\tfrac{1}{8}\sigma^{(1)}_{z}\sigma^{(4)}_{z}\right)} H_1 ~T_\text{I},
\end{align}
or, in the general-$N$ case,
\begin{align}
\label{QFTNseries}
T_\text{F} ~H_\Nd~ P_{\Nd-1} ~H_{\Nd-1}~ \ldots P_{2} ~H_{2}~ P_{1} ~H_{1} ~T_\text{I},
\end{align}
where
\be
\label{seriesU}
P_n=e^{i\tfrac{\pi}{4}\sum_{l=n+1}^\Nd 2^{n-l}\sigma^{(n)}_{z}\sigma^{(l)}_{z}}.
\ee
We will refer to Eq. \eqref{QFTNseries} as the consecutive sequence.
Here $H_k$ is the Hadamard gate, applied to qubit $k$, and $T_\text{I}$ and $T_\text{F}$ are phase gates, applied to all $\Nd$ qubits at the beginning and at the end of the circuit, $T_\text{I} = \prod_{k=1}^\Nd T_k(\phi_{\text{I},k})$, $T_\text{F} = \prod_{k=1}^\Nd T_k(\phi_{\text{F},k})$,
with $\phi_{\text{I},k}=\tfrac{\pi}{4}(1-2^{-k+1})$ and $\phi_{\text{F},k}=\tfrac{\pi}{4}(1-2^{k-\Nd})$.
In a more compact form we have
\bse
\begin{align}
&T_\text{I} = e^{i \sum_{k=1}^\Nd \tfrac{\pi}{4}(1-2^{k-\Nd}) \sigma^{(k)}_{z}},\\
&T_\text{F} = e^{i \sum_{k=1}^\Nd \tfrac{\pi}{4}(1-2^{-k+1}) \sigma^{(k)}_{z}}.
\end{align}
\ese
Note that a phase gate represents a mere shift of the phase of the driving field, rather than a physical modification of the qubit. As such, it is not associated to any infidelity.

In the rest of this section we show that by a proper tailoring of the coupling matrix $J$ it is possible to execute simultaneously the whole sequence $P_k$ (cf. Eq. \eqref{seriesU}) of conditional gates enclosed between two Hadamard gates (cf. Fig. \ref{fig1}). The tailored evolution operator $U$ we denote as $U(J)$ in order to highlight its dependence on the coupling matrix $J$, which will different for different $P_n$. In units where $t=1$, $U(J)$ is defined completely by the matrix $J$.
Now in place of the sequence \eqref{QFTNseries} we use the following
\be
\label{QFTNseries2}
T_\text{F} ~H_\Nd~ U(J^{(\Nd-1)}) ~H_{\Nd-1}~ \ldots U(J^{(2)}) ~H_{2}~ U(J^{(1)}) ~H_{1} ~T_\text{I},
\ee
which we call parallel sequence.

Below we will determine the elements $J^{(n)}_{kl}$. For conciseness we consider the implementation with four ions ($\Nd=4$), shown in Fig. \ref{fig1} (middle).
$J^{(n)}_{kl}$ must satisfy the phase relations listed in Eq. \eqref{phases}:
\be
\label{sys1}
J^{(k)}_{kl} = \frac{\pi}{2^{l-k+1}},
\ee
where $k=1,\ldots, \Nd-1$ and $k+1\leq l\leq \Nd$ with $\Nd=4$.
Note that no net phase shift must take place between ions 2, 3 and 4 in the application of $U(J^{(1)})$ and 3 and 4 in the application of $U(J^{(2)})$.
This leads to
\bse
\label{sys2}
\begin{align}
    J^{(1)}_{23} = J^{(1)}_{24}  & = 0,    \\
    J^{(1)}_{34} + J^{(2)}_{34}  & = 0.  \label{sys2c}
\end{align}
\ese
In Eq. \eqref{sys2c} we account for the fact that $U_{34}(J^{(1)})$ and $U_{34}(J^{(2)})$ commute with the Hadamard gate on ion 2. Therefore they can be combined into a single gate $U_{34}(J^{(1)}+J^{(2)})$, which must produce no net effect.

It is important to note that the righthand side of Eqs. \eqref{sys1} and \eqref{sys2} is defined up to the addition of $2\pi z$, where $z$ is an arbitrary integer, $z \in \mathbb{Z}$.
This offers larger flexibility for the values of $J$.

Now we assume that $J^{(n)}$ can be factorized, $J^{(n)}_{kl} = \pi \alpha a^{(n)}_k a^{(n)}_l$,
where we introduce the physical parameter $\alpha$ in order to obtain dimensionless coefficients $a^{(n)}_k$.
Then a solution to Eqs. \eqref{sys1} and \eqref{sys2} for any $\Nd$ is
\bse
\label{factored}
\begin{align}
    a^{(n)}_{k}   & = 0 & \text{for } k<n,\\
    a^{(n)}_{k}   & = \frac{1}{\sqrt{(2 q_n - q_{n-1})2^{N-n+3}}} & \text{for } k=n,    \\
    a^{(n)}_{k}   & = \frac{2^{-k+n-1} +2 c_k^{(n)}}{a_n^{(n)}} & \text{for } k>n,
\end{align}
\ese
where $q_k$ and $c^{(n)}_{k}$ are arbitrary integers, $2q_k\neq q_{k-1}$ and $q_0=0$. A simple example is when $q_k=1$ ($k>0$) and $c^{(n)}_{k} = 0$.

\section{Physical realization with ion traps}

Below we illustrate our method with homogeneous magnetic-gradient ion traps \cite{Wunderlich},
where the implementation of the $U$-gate is appealingly simple: it occurs naturally in the time evolution of the energy eigenstates of the Hamiltonian \eqref{Hamiltonian}.

\subsection{Hamiltonian of the system and implementation of the gate set}\label{Sec:realizationA}

Consider a system of $\Nd$ ions, confined strongly in two orthogonal directions ($x$ and $y$) and
by a weaker potential along a third direction ($z$), known as the trap axis.
The ions are laser cooled so that the ions
form a linear chain. Two internal levels of each ion,
which we denote as $\ups$ and $\downs$, form a qubit. These can be a pair of hyperfine ground states, serving as effective spin-1/2 states.
In addition, a magnetic field $\vec{B}=B(z)\vec{e}_z$ is applied, 
such that the ions experience a gradient along the trap axis.

\subsubsection{Hamiltonian of the system}
The Hamiltonian of the system of $\Nd$ ions, up to second order in the ions' vibrational motion, is given by
\be
\label{Hamiltonian2}
\HS=\frac{\hbar}{2}\sum_{n=1}^\Nd \omega_n\sigma^{(n)}_{z} +
\sum_{n=1}^\Nd \hbar\nu_n a_n^{\dagger}a_n - \frac{\hbar}{2}\sum_{n<m=1}^{\Nd}J_{nm}\sigma^{(n)}_{z}\sigma^{(m)}_{z}.
\ee
The first term is the sum of the internal energies of the two-level ions located at equilibrium positions $z_n$ and represented by Pauli's matrices $\sigma^{(n)}_{z}$. Due to the magnetic field gradient, the ions' transition resonant frequencies are position-dependent, $w_n = g\mu_\text{B} B(z_n)/\hbar$, where $g$ and $\mu_\text{B}$ are respectively the electron's g-factor and the Bohr magneton. The second term describes $\Nd$ axial collective vibrational modes with frequencies $\nu_n$; $a_n^{\dagger}$ and $a_n$ are respectively the creation and annihilation operators of mode $n$. The last term realizes a long-range pairwise interaction between the spins, known as spin-spin coupling \cite{Wunderlich}, whereby spin $n$ is coupled to spin $m$ with strength
\be
\label{Jnm}
J_{nm}=\frac{\left(g\mu_\text{B}\right)^2}{2\hbar}\left.\frac{\partial B(z)}{\partial z}\right|_{z_{0,n}}\left.\frac{\partial B(z)}{\partial z}\right|_{z_{0,m}}(\hes^{-1})_{nm}.
\ee
Here $\hes_{mn}$ is the Hessian of the total potential energy, which is the matrix of second-order partial derivatives of the external plus Coulomb potential taken at the equilibrium positions $z_{n}$.
Thus the spin-spin coupling is defined by the values of the magnetic field gradient and the shape of the potential surface around the equilibrium positions of the ions.
In the following we will discover that only the third term is essential for the implementation of QFT.

For the complete description of the system, we introduce the density operator $\r(t)$, which in the absence of decoherence, as assumed below, obeys the Liouville equation $d\r(t)/d t=-i \left[\HS(t),\r(t)\right]$, with $\HS(t)$ being the Hamiltonian. The formal solution is $\r(t)=\U(t)\r(0)\U^\dagger(t)$, where $\U(t)$ is the propagator (the evolution matrix). For our time-independent Hamiltonian \eqref{Hamiltonian2} we have $\U(t)=\exp(-i \HS t/\hbar)$.

We assume that initially the system resides in a pure state, which is a product of a spin part and a motional part,
$\r(0)=\r_{\text{internal}}(0)\otimes \r_{\text{motional}}(0)$.
This happens, for example, when the spins are prepared in a particular state as part of the initialization of the system.  Because the interactions that we will consider below do not couple the vibrational motion to the spin states, the latter are evolved independently of the motional states and thus $\r(t)$ preserves its product form over time. Therefore with $\U(t)$ we will refer only to the spin-driving part of the propagator, as we will be only interested in the dynamics of the logical basis states. For this reason we will omit the second term from the Hamiltonian \eqref{Hamiltonian2}. Moreover, because $\U(t)$ is independent of the vibrational state and it does not matter whether the vibrational state is cooled to the ground state or not.

\subsubsection{Implementation of the gate set}

\begin{figure}[tb]
\begin{center}
\includegraphics[angle=0,width=0.55\columnwidth]{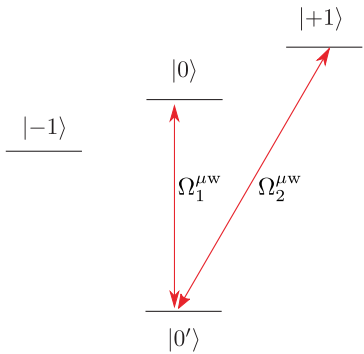}
\end{center}
\caption{Two bases of quantum states are used. The coupled basis comprises $\ket{0'}$ and the magnetic field sensitive state \mbox{$\ket{+1}$}. The uncoupled basis encodes the qubits and is formed by magnetic field insensitive states $\ket{0}$ and $\ket{0^\prime}$, which belong to different hyperfine manifolds. An ion is brought to the uncoupled basis to suppress its interaction with the other ions.}
\label{bases}
\end{figure}

In our implementation we will use two bases, shown in Fig. \ref{bases}. The \emph{coupled} basis is formed by the magnetic-field sensitive states $\ket{+1}$ and $\ket{0^\prime}$. Ions residing in this basis are coupled by the magnetic-field gradient and thus can participate in conditional gates. The \emph{uncoupled} basis is formed by the magnetic-field insensitive states $\ket{0}$ and $\ket{0^\prime}$. We will uncouple ions when we want to suppress interaction with the others. Below we describe a possible scheme to implement coupling and uncoupling.

To illustrate the scheme we consider the state vector of the three states $\ket{\psi} = (c_{0'}, c_{+1}, c_{0})^\text{T}$.
The general initial (coupled state) is $\ket{\psi_\text{I}} = (\cos\theta, e^{i\phi}\sin\theta, 0)^\text{T}$.
We implement a $\pi$-pulse on the transition $\ket{0}_n\leftrightarrow\ket{0'}_n$,
\be
\Pi_{0',0}=\left[
    \begin{array}{ccc}
      0 & 0 & 1 \\
      0 & 1 & 0 \\
      1 & 0 & 0 \\
    \end{array}
  \right],
\ee
and a $\pi$-pulse on the transition $\ket{+1}_n\leftrightarrow\ket{0'}_n$,
\be
\Pi_{0',+1}=\left[
    \begin{array}{ccc}
      0 & 1 & 0 \\
      1 & 0 & 0 \\
      0 & 0 & 1 \\
    \end{array}
  \right].
\ee
Thus we obtain
\be
\ket{\psi}_{\rm uncoupled} = \Pi_{0',0} \, \Pi_{0',+1} \, \Pi_{0',0} \ket{\psi}_{\rm coupled}.
\ee

The second $\Pi_{0',0}$-pulse automatically takes care of a potential problem that the $\Pi_{0',0}$-transition is Zeeman insensitive and thus swaps populations for all ions. It can be omitted, for speed-up and in the sense of 'fewer gates=lower errors', but then some book-keeping is required, where all the populations are. If applied, a useful improvement is then to do the second rotation $\ket{0}_n\leftrightarrow\ket{0'}_n$ with a phase of $\pi$, so pulse length errors would be compensated as the Rabi frequencies for all ions on that transition might not be exactly identical.

{\it The $U$-gate}. In a frame rotating with the spins' transition frequencies $w_n$, our system is described by the Hamiltonian $\HS^\prime = F^\dagger \HS F-i \hbar F^\dagger \dot{F}$, where $F=\exp(-i t/2\sum_{n=1}^{\Nd}\omega_n \sigma^{(n)}_{z})$ \cite{note}. The transformed Hamiltonian $\HS^\prime$ contains only the spin-spin coupling term and thus it coincides with the Hamiltonian \eqref{Hamiltonian}. We remind that so far no external interaction is applied to the ion chain other than the trapping electric potentials and the magnetic gradient field. Hence the entangling gate $U_{kl}(\phi)$ (cf. Eq. \eqref{U}) occurs from free evolution of ions $k$ and $l$ for time $t_{kl}=2\phi/J_{kl}$ with all other ions being uncoupled.

{\it The $R$-gate}. The one-qubit rotation $R_k(\theta,\phi)$ is obtained using a microwave (mw) pulse resonant with the spin-flip transition frequency of ion $k$.
The mw pulse delivers an oscillating magnetic field in the $x$-$y$ plane
\be
\vec{\Bosc}=\Bosc\left[\cos \left(\omega t+\phi\right) ~\vec{e}_x + \sin \left(\omega t+\phi\right) ~\vec{e}_y\right].
\ee
It couples to the spin thereby driving the rotation
\be
\label{UA}
R_k(\theta,\phi)=\exp[-\tfrac12 i \theta \left(\sigma^{(k)}_x \cos\phi+\sigma^{(k)}_y \sin\phi \right)],
\ee
where $\theta=\Omega t$ is the pulse area and $\Omega=g\mu_\text{B}\Bosc/\hbar$ is the coupling between the spin and the driving field.
The angle $\phi$ is set by the phase of this field.

{\it The $H$-gate}. The Hadamard gate, up to an unimportant phase, is obtained with two pulses,
\be
\label{Hadamard}
H=R(\pi/2,-\pi/2)R(\pi,0),
\ee
where the second pulse is phased with $-\pi/2$ relative to the first.

Armed with a gate set, in the following we show how to realize QFT.

\subsection{Realization of the Fourier transform}

\subsubsection{The consecutive sequence}
For brevity, we illustrate the implementation with four ions. The gate sequence is shown in Eq. \eqref{QFT4series} and is depicted in Fig. \ref{fig1} (middle). After the application of the $T$-gates we uncouple all ions and follow the steps:
i) Couple ion 1 and then apply the Hadamard gate $H$ to it. Next continue with the implementation of the series $P_1$.
ii) Couple ion 2. Now ions 1 and 2 can interact. Wait for time $T=(\pi/4)/J_{12}$ thereby applying the gate $U_{12}(\pi/4)$ and then uncouple ion 2. Continue likewise with the gates $U_{13}$ and $U_{14}$.
iii) After completing $P_1$ uncouple ion 1. Thus it is no longer manipulated until the application of the final phase gate (cf. Fig. \ref{fig1} (middle)). Steps i)-ii) complete the encircled part of the circuit. Likewise proceed with the rest.
In the general-$N$ case one follows Eq. \eqref{QFTNseries}.

\subsubsection{The parallel sequence}
We now discuss the sequence \eqref{QFTNseries2} depicted in Fig. \ref{fig1} (bottom), where we focus on the realization of the tailored operator $U(J)$. This operator can be achieved using oscillating laser or magnetic fields, for example. Below we employ magnetic fields, while we note that with laser fields the implementation is essentially the same.

Consider $\Nd$ trapped ions, as described above, which now interact with a bichromatic magnetic field with frequencies $\omega_r=\omega_{0,n}-\omega_\cm+\delta$ (``red'') and $\omega_b=\omega_{0,n}+\omega_\cm-\delta$ (``blue'') ($\omega_\cm\gg\left|\delta\right|$) tuned near the centre-of-mass (c.m.) mode $\omega_\cm$ with detuning $\delta$ ($\omega_\cm\gg \left|\delta\right|$). Here $\omega_{0,n}$ is the atomic transition frequency for ion $n$. The interaction Hamiltonian in the Lamb-Dicke limit and in the rotating-wave approximation is given by \cite{Monroe,Wunderlich2001,Ivanov2015} ($\hbar=1$)
\be
H_{\rm I} = \sum_{k=1}^{N}g_{k}\sigma(\varphi_{k}^{+})(a^{\dag}e^{i\delta t-i\varphi_{k}^{-}}+ ae^{-i\delta t+i\varphi_{k}^{-}}).\label{Hbich}
\ee
Here $g_{k}$ is the individual (time-independent) spin-phonon coupling for ion $k$, $a^{\dag}$ and $a$ correspond to the {\rm{c.m.}} vibrational mode and $\sigma(\varphi_{k}^{+})=e^{-i\varphi_{k}^{+}}\sigma_{k}^{+}+e^{i\varphi_{k}^{+}}\sigma_{k}^{-}$ with $\sigma_{k}^{\pm}$ being the spin raising and lowering operators for ion $k$.
The spin and motional phases are defined as $\varphi_{k}^{+}=\frac{1}{2}(\varphi_{k}^{\rm b}+\varphi_{k}^{\rm r})$ and $\varphi_{k}^{-}=\frac{1}{2}(\varphi_{k}^{\rm b}-\varphi_{k}^{\rm r})$, where the phases $\varphi_{k}^{\rm b}$ and $\varphi_{k}^{\rm r}$ correspond to the blue and the red component of the oscillating magnetic field. Hereafter we assume that the blue and the red B-field are counterpropagating, which implies that $\varphi_{k}^{+}=0$ and $\varphi_{k}^{-}=\varphi_{k}$, such that $\sigma(\varphi_{k}^{+})=\sigma_{k}^{x}$.

The propagator is obtained using the Magnus expansion \cite{Magnus}:
\be
U = D(\alpha)\exp\left(i\frac{\delta t-\sin\delta t}{\delta^2}\tsum_{k<p=1}^{\Nd} 2g_k g_p \sigma_k^x \sigma_p^x \right),
\ee
where $D(\alpha)=e^{\alpha a^{\dag}-\alpha^{\dag}a}$ and $\alpha=(1-e^{i\delta t})/\delta\sum_{k=1}^{\Nd}g_k\sigma_k^x$.
A more detailed study of interaction with bichromatic fields can be found in Ref. \cite{Monroe}.
At time $\tau = 2\pi/\delta$ the displacement $D(\alpha)$ vanishes and we obtain
\be
U = \exp\left(\frac{2\pi i}{\delta^2}\sum_{k=1}^\Nd g_k^2\right)\exp\left(\frac{4\pi i}{\delta^2} \sum_{k<p=1}^{\Nd}g_k g_p \sigma_k^x \sigma_p^x\right).
\ee
This propagator has the form as shown in Eq. \eqref{propagator} with $J_{kl}\propto g_k g_l$.
The global phase factor can be easily compensated for at the end of the circuit.

\subsection{Example with $N=3$ qubits}
Here we specify our implementation \eqref{QFTNseries} for a system of three qubits. Following our proposal (cf. Eq. \eqref{QFTNseries}), the three-qubit Fourier transform is obtained with
\begin{align}
\label{QFT3}
T_\text{F} H_3 e^{i \pi\left(\tfrac{1}{8}\sigma^z_2\sigma^z_3\right)} H_2 e^{i \pi\left(\tfrac{1}{8}\sigma^z_1\sigma^z_2+\tfrac{1}{16}\sigma^z_1\sigma^z_3\right)} H_1 T_\text{I}.
\end{align}
We can rearrange this sequence so that we can employ simultaneous interactions in the first two conditional gates and thereby gain some speed-up. The rearranged sequence is
\bea
\label{series3}
R_2(A_{2},\tfrac{3\pi}{4}) R_3(\tfrac{\pi}{2},-\tfrac{\pi}{2}) U_{23}(T_3) R_1(\pi,\tfrac{3\pi}{16}) R_2(A_1,\tfrac{3\pi}{4})  \\
R_3(\pi,-\tfrac{3\pi}{16}) U(T_2) R_3(\pi) U(T_1) R_1(\pi) R_2(\pi)R_3(\pi) H_1, \notag
\eea
The derivation of this circuit is shown in Ref. \cite{Experiment}. Here we have used Eq. \eqref{Hadamard} and
we have
\bse
\bea
T_1 &= \frac{\pi}{8}\left(\frac{1}{J_{12}} + \frac{1}{2J_{13}}\right), \\
T_2 &= \frac{\pi}{8}\left(\frac{1}{J_{12}} - \frac{1}{2J_{13}}\right).
\eea
\ese
The duration $T_3$ and the pulse areas $A_1$ and $A_2$ are obtained from the following set of equations
\bse
\bea
\tfrac{1}{\sqrt{2}} e^{i\frac{\pi}{16}(\alpha+2)} e^{i J_{23}T_3/2} - &\sin\tfrac{A_1}{2} \sin\tfrac{A_2}{2} e^{i J_{23}T_3} \notag \\
+ &\cos\tfrac{A_1}{2} \cos\tfrac{A_2}{2} = 0,\\
\tfrac{1}{\sqrt{2}} e^{i\frac{\pi}{16}(\alpha-2)} e^{i J_{23}T_3/2} - &\sin\tfrac{A_1}{2} \cos\tfrac{A_2}{2} e^{i J_{23}T_3} \notag \\
- &\cos\tfrac{A_1}{2} \sin\tfrac{A_2}{2} = 0.
\eea
\ese
where $\alpha=J_{23}/J_{13}$.
Note that the sequence \eqref{series3} is tunable in the sense that it is valid for \textit{any} coupling matrix $J$.

For homogeneous traps having nearly harmonic electrostatic potential and magnetic field gradient $b=20$ T/m we have $T_1\approx 3.241$ ms, $T_2\approx 0.558$ ms and $T_3\approx 4.478$ ms.
Due to the simultaneous interactions, the total duration of the algorithm is now $T_\text{tot}=T_1+T_2+T_3\approx 8.3$ ms. For comparison, the duration using sequence $\eqref{QFTNseries}$ is $\pi/(4J_{12})+\pi/(8J_{13})+\pi/(4J_{23})\approx 11.2$ ms. For the pulse areas we have $A_1\approx 0.654\pi$ and $A_2\approx 0.771\pi$.

\section{Discussion and conclusions}

The total implementation time for the consecutive sequence from Eq. \eqref{QFTNseries} is estimated by summing up the time it takes to produce each $U$-gate,
\be
T(\Nd)=\pi \sum_{k=1}^{\Nd - 1}\sum_{l=k + 1}^\Nd\frac{2^{-(l - k + 1)}}{J_{kl}}.
\ee

We obtain $T(\Nd)\approx c \Nd^2$, 
where, for example, $c$ is on the order of Milliseconds for experiments with $^{171}$Yb$^+$, gradients of $b\approx 20$ T/m and an axial trap frequency of $\nu_0 \approx 2\pi \cdot 200$~kHz. The implementation needs 2($\Nd$-1) phase gates, $N$ Hadamard gates and $O(\Nd^2)$ $\pi$-pulses for coupling and uncoupling ions, amounting to O($\Nd^2$) pulses overall.

The time duration of the parallel sequence \eqref{QFTNseries2} where now the conditional interactions run simultaneously is given by
\be
T(\Nd)=\frac{\pi}{4}\sum_{n=1}^{\Nd-1}\frac{1}{J^{(n)}_{n,n+1}}.
\ee
Using Eq. \eqref{factored} we obtain $T(\Nd) = \frac{\Nd-1}{4\alpha} \sim \Nd$. Note that a quadratic speed-up is achieved.

\acknowledgments
This work has been supported by the European Community's Seventh Framework Programme (FP7/2007-2013) under Grant Agreement No. 270843 (iQIT)


\begin{thebibliography}{99}
\bibitem{note}
The additional phases which will occur in the rotating frame can be easily undone by phase gates at the end of the circuit.
\end{thebibliography}
\end{document}